\documentclass[longbibliography]{revtex4-1}
\usepackage{verbatim}
\usepackage{listings}
\usepackage{mathtools}
\usepackage{lipsum}
\usepackage{graphicx}
\usepackage[outdir=./]{epstopdf}

\begin{document}
\bibliographystyle{plain}
\title{Modelling curvature of a bent paper leaf}
\author{Sasikanth Raghava Goteti}
\email{raghavas@thoughtworks.com}
\homepage{https://sasikanth3.github.io/}
\affiliation{Thoughtworks ,India}
\date{\today}
\keywords{Machine Learning, Algebraic Geometry, Elastica, Euler Spiral}
\begin{abstract}
 In this article, we briefly describe various tools and approaches that algebraic geometry has to offer to straighten bent objects.
 throughout this article we will consider a specific example of a bent or curved piece of paper which in our case acts very much like an     
 elastica curve. We conclude this article
 with a suggestion to algebraic geometry as a viable and fast performance alternative of neural networks in vision and
 machine learning.The purpose of this article is not to build a full blown framework but to show possibility of using 
 algebraic geometry as an alternative to neural networks for recognizing or extracting features on manifolds .
\end{abstract}
\maketitle

\section{Introduction}
Elliptic curves and curves from algebraic geometry have surfaced many times in the recent years in fields related to computer vision ~\cite{Mumford} and machine learning. Mumford has successfully demonstrated that elastica curves can be represented as the curve that maximizes the likelihood of a random curvature variable.Levin has made a detailed historical report of these curves in detail ~\cite{Levien} .While most of the applications have been indirect in nature in this article we tried to find a direct application to
elastica curves and Euler spirals.In computer vision modifying a scanned document to a shape that is geometrically flat is a sort out problem. Very effective solutions already exist that can de-skew ,rotate and linearize documents. While most of them are supervised approaches that require large data sets before we can straighten the documents.In this article we try to build a solution that strictly involves geometric modeling of the surface of a paper without requiring to use any supervised machine learning techniques.These applications can be very useful in flattening the surface of a scanned document attached to a book using a hand-held device like mobile phone. This approach can be used very effectively to convert the curved graphics on the center fold end of a book which are often very difficult to read from a scanned pdf file. Mobile applications like office lens and cam scanner can use this application to make the scanned view more readable without putting too much stress on resource crunched mobile phones.

\section{Problem Statement}
When a book is opened there is a certain curvature that surface of the pages forms, the corners of the pages are usually curved and is often difficult to read the inner parts of .It becomes even harder to read the pages when it is a document scanned from a book with lot of pages. The curvature of the document is shown in a side view in figure 1. The problem is to identify the curve in figure 1 and use computational techniques to geometrically flatten the scanned documents .
\begin{figure*}
  \centering
  \caption{OpenBook}
  \includegraphics[width=0.8\textwidth]{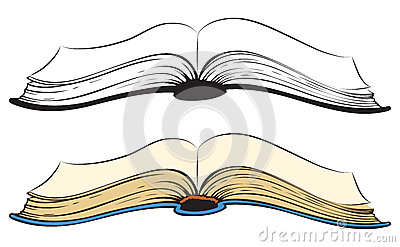} 
\end{figure*}

\section{Proposed Solution}
The bending of a paper material is similar to that of a bending beam. Historically there are two approaches for resolving this problem using calculus of variation . Minimizing the square of infinitesimal curvature leads to elastica solution or MEC and in case of minimizing curvature differential we arrive at MVC . At small angles of deflection both the solutions are equivalent. MVC approach is computationally and mathematically easy to represent and for the most part we would use the Cornu spiral due to its computational ease and the natural spiral nature that it provides for the problem.The two curves can be parametrically plotted in sage as below

\begin{lstlisting}
s,x,y,p = var('s x y p')
e = var('e', latex_name=r'\vartheta')
k=0
p=2.4
l=7*pi/6
f = s**2
\end{lstlisting}
\[
x=\sin(s^2)
\]
\[
y=\cos(s^2)
\]
\begin{figure*}
\centering
\includegraphics{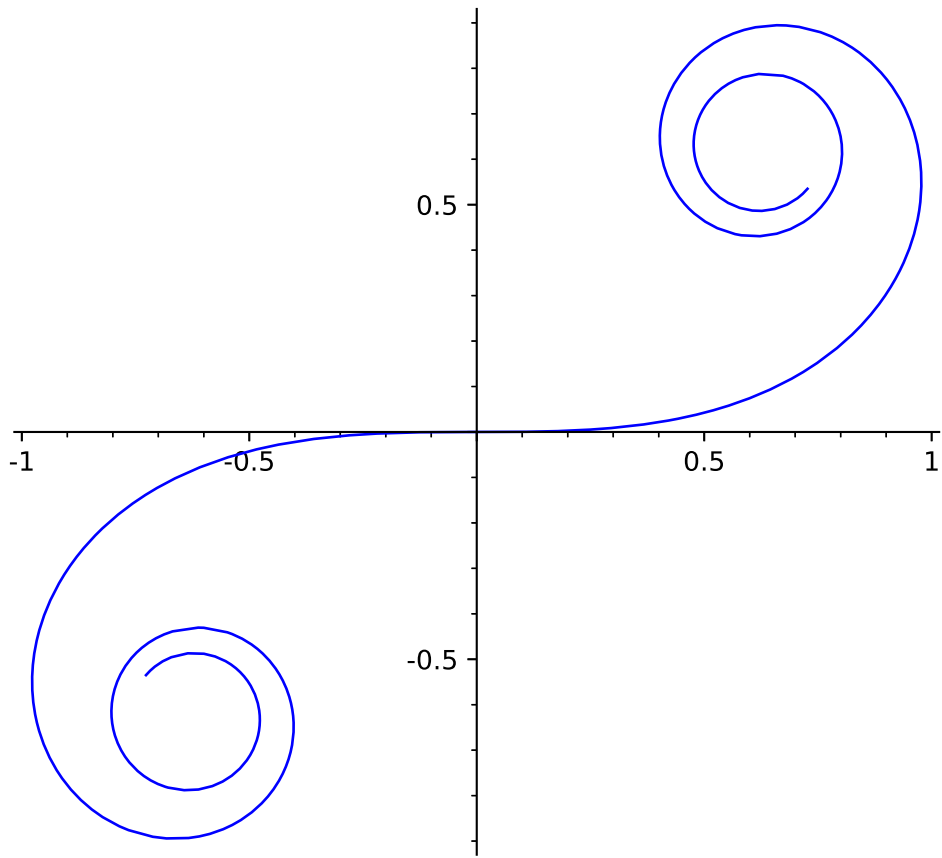}
\caption{Cornu Spiral}
\end{figure*}

A similar bending plot for smaller angles can be generated for using elastica based models it involves Jacobi integrals and a sage math based code and plot for it would be:

\begin{lstlisting}
k=0.3
parametric_plot((lambda s: 2*k*jacobi('cn', s, k),lambda s: 2*elliptic_eu(s,k)-s ), (0, 1))
\end{lstlisting}

\begin{figure*}
\centering
\includegraphics{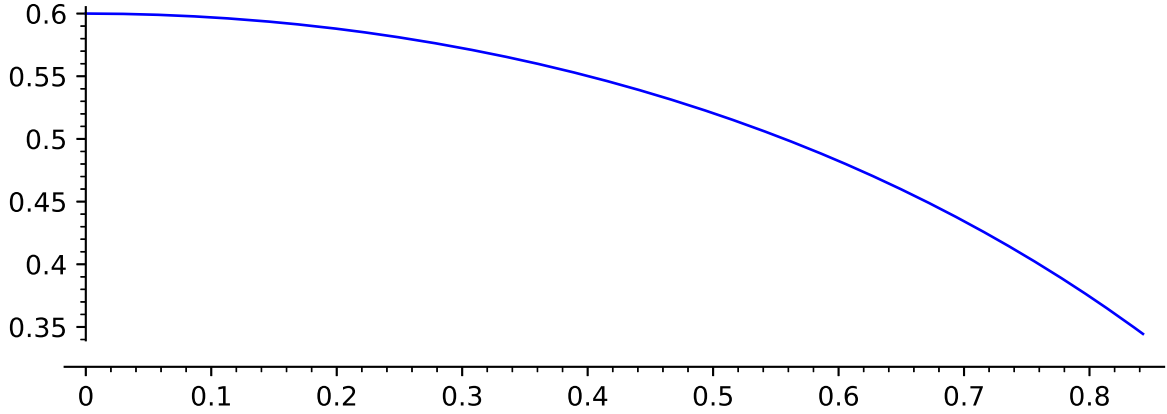}
\caption{Elastica bent}
\end{figure*}

\section{Modeling The cross section of a page}

Euler's assumption for modeling the curvature of minimizing variations is very clearly described in his paper where he assumes the curvature of a bent rod or elastica is directly proportional to the length of the arc. which eventually leads to a Cornu spiral , but the main difference is, a typical Cornu spiral has its coordinates placed at the non-axial end of the cantilever loading. So we can modify Euler's approach by placing coordinates at the axial ending

In Euler's coordinate system

$\kappa=s$

$\frac{d\theta}{ds} =s$

if the cross length of the paper is l the same above equation cab be written as

$\kappa=s$

$\frac{d\theta}{ds} =(m-s)$

which would lead to the parametric equations 

\[
x=\sin(ms-(s^2)/2)
\]
\[
y=\cos(ms-(s^2)/2)
\]

a better way to handle this would be to actually translate the Cornu spiral to the end using brute force transformation as that would give us the advantage of bending the paper even in negative angles, a final plot for the equations would look like these:

\begin{figure*}
\centering
\includegraphics{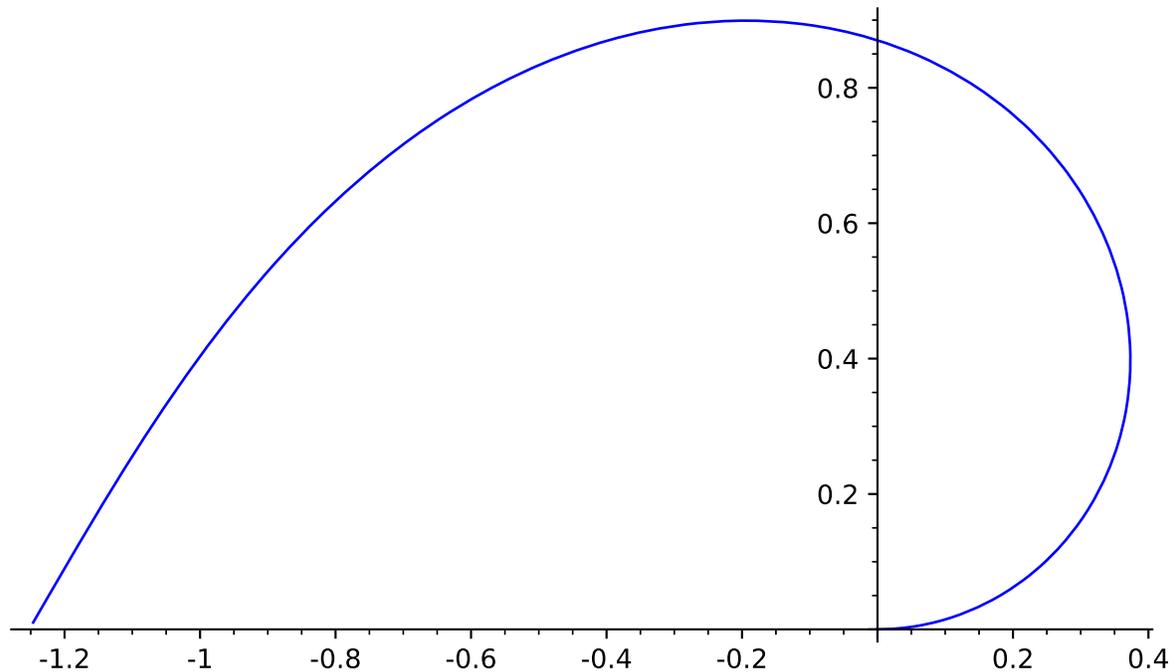}
\caption{Paper bent fully}
\end{figure*}

The above plot is a very easy way to represent an axial bending of a beam or a paper but in reality there is always an axial force that is applied while bending a paper which give it the more flattened shape.

\section{Stretching the elastica action}
When you laterally stretch the elastica the new parametric coordinates can be expressed as functions of old x,y . This is particularly useful to note that all curvatures of elastica can be represented with in one degree of freedom using the arc length.

so if we call $x^{'}$ and $y^{'}$ as the new parametric coordinates one can represent the equation in terms of old coordinates and arc length $s$ as $x^{'} =  X(x,s)$ and $y^{'} = Y(y,s)$ it is easy and convenient to see that X and Y are same as lateral action of force is independent of and chosen coordinate system  we can replace $X$ and $Y$ by a new action function $f$. which acts over the transformation coordinates $x^{i}$ which in this case are $x$ or $y$

\newtheorem{thm}{Theorem}
\begin{thm}[Theorem]
The stretch action function is directly proportional to the coordinate and also the arc length
\end{thm}

let us rewrite the action function as $F(x^{i},s)$ a Taylor series expansion first in one coordinate $x^{i}$ and then in $s$
it is easy to see that the transformation function easily vanishes when $x^{i}=0$ or $s=0$ or $\theta=0$
hence the constant term is going to vanish in Taylor expansion as $F(0,s)=0$ and $F(x^{i},0)=0$ ,so x and s can be extracted out.
and hence $F(x^{i},s) = xsG(x^{i},s)$

For the ease of our computation we can actually re write the above parametric equations in lieu with the above theorem and rewrite them as

 $x^{0}= s^{\lambda} (\int_{0}^{s} cos(m*s-(s**2)/2) ds)$ 
 $x^{1}= s^{\lambda} (\int_{0}^{s} sin(m*s-(s**2)/2) ds)$ 

 here lambda is a stretching constant(the stretching parameter can actually be replaced with any function of $s$ but to keep the number of parameters as little as possible we stick with just one constant)

 by using the approximations we can re-plot the curve with a modeled lateral force

\begin{lstlisting}
#final cornu based model
s,x= var('s x')
assume(x>0)
l= 2.170803
e= -0.78622
f= s**2
m= numerical_integral(cos(f),0,l)[0]
n= numerical_integral(sin(f),0,l)[0]
a= numerical_integral(cos(f),0,e)[0]
b= numerical_integral(sin(f),0,e)[0]
theta= arctan((b-n)/(a-m))
c= cos(theta)
s= sin(theta)
def x(t): return c*(numerical_integral(cos(f),0,t)[0] -m) + s*( numerical_integral(sin(f),0,t)[0] -n )
def y(t): return s*(numerical_integral(cos(f),0,t)[0] -m) - c*(numerical_integral(sin(f),0,t)[0] -n )
parametric_plot((lambda z : x(z)*(l-z)**2,lambda z: y(z)*(l-z)**2), (e, l))
\end{lstlisting}

\begin{figure*}
\centering
\includegraphics{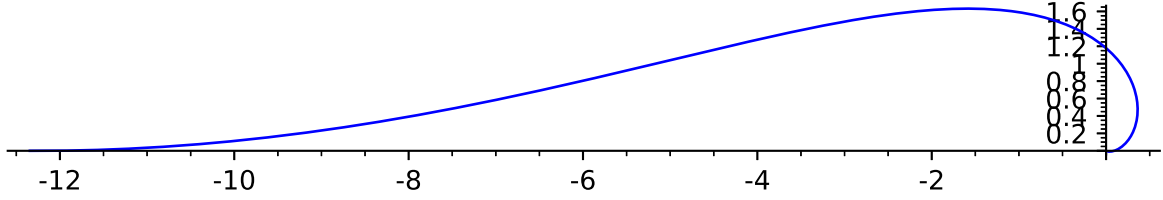}
\caption{Paper Bent And Laterally Stretched}
\end{figure*}

\section{Computational model}
We can use the above model to build a complete curvature equation of a paper or elastica as inscriptions on a manifold.
Then the problem of straightening the leaves on a book would just be the inverse of the function $x^{'}(s)$
i.e $F^{-1}(x^{'})*x^{'}$ is a flattened model of leaves on the book.An easier solution to this problem is to actually create a look up table and create linear array for all the values let $X$ be the matrix that transforms a flat paper to a curved one then the inversion of it could be solved computationally on an image which $A_{mXn}$ as multiplication of column matrix

\[A_{mXn} \left( \begin{array}{ccc}
F^{-1}(x^{'}) \\
F^{-1}(x^{'}) \\
.\\
.\\
.\\
F^{-1}(x^{'}) \end{array} \right) = X\] 

\section{Future Opportunities}
This problem leads us to model various other curves that we see in our day to day life for example facial recognition can be heavily dependent on various types of elastica and spiral based splines as much of human flesh and tissues acts like an elastica  giving it the elastica type curvature cheeks and other human features.We can think of algebraic geometry as a very good alternative to neural networks for learning features of various curves and vision based problems.Unlike neural networks which requires lot of statistical data to identify  features algebraic geometry proposes a more natural human mind like approach of extracting features one or two samples of data are all that are probably required to completely understand various curves / features that are present on a human face

\end{document}